\documentclass{Interspeech2024}
\usepackage{booktabs}
\usepackage{multicol, multirow}
\usepackage{amsmath,pifont}
\usepackage{caption}
\usepackage{url}
\usepackage{xcolor}
\usepackage{subcaption}
\usepackage[inline]{enumitem}
\usepackage{tablefootnote}
\newcommand{\xmark}{\ding{55}}%
\definecolor{xiaomi_gray}{HTML}{A9A9A9}
\usepackage{cleveref}

\interspeechcameraready

\title{Streaming Audio Transformers for Online Audio Tagging}

\address{
 AI Lab, Xiaomi Corporation, China}
\email{\{dinkelheinrich,yanzhiyong, wangyongqing3, zhangjunbo1, wangyujun,wangbin11\}@xiaomi.com }

\name{Heinrich}{Dinkel}
\name{Zhiyong}{Yan}
\name{Yongqing}{Wang}
\name{Junbo}{Zhang}
\name{Yujun}{Wang}
\name{Bin}{Wang}


\begin{document}

\maketitle

\begin{abstract}
    
Transformers have emerged as a prominent model framework for audio tagging (AT), boasting state-of-the-art (SOTA) performance on the widely-used Audioset dataset. 
However, their impressive performance often comes at the cost of high memory usage, slow inference speed, and considerable model delay, rendering them impractical for real-world AT applications. 
In this study, we introduce streaming audio transformers (SAT) that combine the vision transformer (ViT) architecture with Transformer-Xl-like chunk processing, enabling efficient processing of long-range audio signals. 
Our proposed SAT is benchmarked against other transformer-based SOTA methods, achieving significant improvements in terms of mean average precision (mAP) at a delay of 2s and 1s, while also exhibiting significantly lower memory usage and computational overhead.
\end{abstract}

\section{Introduction}

Audio tagging (AT) is a task that aims to label specific audio content into a fixed set of sound event classes, e.g., dog barking or people speaking.
Applications of AT systems include aid for the hearing impaired, smart cities and homes~\cite{Bello2018} and general monitoring of sounds~\cite{xiong_construction_at}.
More recently AT systems have found applications on smartphones and smart speakers as a hearing aid for the needy.
The transformer model, originally introduced in~\cite{vaswani2017attention}, which uses self-attention as its core building block, has become a popular method to achieve excellent performance for AT, however, the deployment of transformer architectures in real-world scenarios has been largely neglected.
Previous works using Vision Transformer (ViT) based models such as in~\cite{gong21b_interspeech,gong2022cmkd,chen2022beats,chen2022hts,chong2022masked,niizumi2022masked,baade22_interspeech,koutini22_interspeech} are optimized towards \textit{offline} usage with a \textit{global} context of 10s.
Unfortunately, this approach results in a model response time (delay) of at least 10s.
In our work, we define delay as the amount of data that a model needs to process before generating an output.
Many transformer architectures in AT suffer from a high memory requirement due to their quadratic self-attention complexity, which depends on the amount of data processed at once (10s).

However, real-world applications are \textit{online}, meaning that a model needs to return results as quickly as possible with a minimal delay while having access to a \textit{limited} context (i.e., 1s).
Although one may easily enable ``online'' inference by recomputing a 10s audio segment every e.g., 1s, this practice is inefficient, particularly when leveraging large transformer-based models~\cite{gong21b_interspeech}.
To address this challenge, streaming inference algorithms aim to compute outputs efficiently without the need for recomputation by leveraging caching of previous results.
This work solely focuses on optimising transformer-based models towards streaming inference, since traditionally used 2-dimensional convolutional neural networks (CNNs) are hard to make streamable~\cite{stefanski2023short}.

We point out three essential prerequisites of AT models for real-world deployment, namely: 
\begin{enumerate*}[label=(\Roman*)]
\item A minimal delay in terms of data necessary to output a label, typically on the order of 1-2 seconds.
\item A small memory footprint and low computational complexity.
\item Robust and reliable performance.
\end{enumerate*}
While there exist many works in literature that tackle the problem of low delay~\cite{dinkel2022pseudo}, reducing memory footprint~\cite{chen2022hts,schmid2022efficient} and improving performance~\cite{kong2020panns,gong21b_interspeech,chen2022beats}, no comprehensive investigation has yet tackled all three issues.
Thus, this work proposes \textbf{s}treamable \textbf{a}udio \textbf{t}ransformers (SAT), aimed at real-world usage of transformers for AT.
Our motivation for this work is twofold.
Firstly, it would improve compatibility between AT models and other audio subfields that are streaming, such as automatic speech recognition~\cite{moritz2020streaming,kannan2019large}, keyword spotting~\cite{rybakov20_interspeech,wang2021wake} and source separation~\cite{wang20z_interspeech,rikhye2022closing}.
Secondly, when deployed on stationary hardware like smart speakers, SAT models could act as an anomaly detector for long reoccurring sounds, differentiating between harmless and potentially harmful events, such as a single beep from a fire alarm versus continuous beeping.
As we empirically demonstrate in this work, standard AT models struggle to continuously predict sound events (\Cref{ssec:quantative}).
Our contributions are:
\begin{enumerate*}[label=(\Roman*)]
    \item We experiment with three standard-sized ViT models (Tiny, Small, Base), and optimize the training baseline for AT, aiming to reduce their memory consumption and decrease their floating-point operations per second (Flops).
    \item Based on those three models, we introduce streamable (SAT) variants, denoted as SAT-T (Tiny), SAT-S (Small) and SAT-B (Base). 
    We compare these models with other transformers in the literature and find significant performance improvements for short-delay inference.
\end{enumerate*}

\section{Vision Transformers for Audio Tagging}

Transformers were first proposed for machine translation in~\cite{vaswani2017attention} and quickly became the state-of-the-art (SOTA) approach within the field of natural language processing (NLP) and later~\cite{Dosovitskiy_ViT} the {Vi}sion {T}ransformer (ViT) has been proposed as an adaption to the computer vision domain.
Then, ViT-based transformers were used in AT, where images were replaced with two-dimensional spectrograms~\cite{gong21b_interspeech,xu2022masked}.
The core idea of the ViT framework is the patchification operation, where an input spectrogram $\mathbf{S} \in \mathbb{R}^{F_{spec} \times T_{spec}}$ is first split into $N$ non-overlapping patches of dimension $d$ via a 2-dimensional convolution with a kernel-size $P$ and stride $P$ as:
\begin{equation}
\label{eq:patchification}
\mathbf{X} = \text{Conv2D}(\mathbf{{S}},P,P) = \{\mathbf{x}_1, \mathbf{x}_2, ..., \mathbf{x}_N\}.
\end{equation}

Then these resulting features (tokens) denoted as $\mathbf{X} \in \mathbb{R}^{d \times N}$ are fed into a transformer model, consisting of a self-attention mechanism (\Cref{eq:self_att}) followed by a feed-forward neural network.
Single-headed self-attention computes a similarity measure between linear transformations of the input $\mathbf{X}$ as:
\begin{align}
\label{eq:self_att}
    \mathbf{Y} = \text{softmax}(\frac{\mathbf{QK}^T}{\sqrt{d}})\mathbf{V},
\end{align}
where $\mathbf{K} \in \mathbb{R}^{N\times d},\mathbf{Q} \in \mathbb{R}^{N\times d}$ and $\mathbf{V} \in \mathbb{R}^{N\times d}$ are the key, query and value matrices obtained by a linear transformation $\mathbf{W}_j \in \mathbb{R}^{d\times d}, j \in \{K,Q,V\}$ of the same input $\mathbf{X} \in \mathbb{R}^{N\times d}$ and $\mathbf{Y} \in \mathbb{R}^{N\times d}$ is the output.

Self-attention scales quadratically for both memory and computation cost in regard to the input sequence length $N$.
For ViT based approaches, the sequence length $N$ consists of time and frequency tokens, thus our goal in this work is to reduce the frequency dimension and restrict the number of time frames accessible to the model.

\section{Streaming Transformers}

Transformer-XL-like architectures address the length limitation by performing a chunk-wise sequence processing.
Specifically, the input $\mathbf{X} \in \mathbb{R}^{d\times N}$ of length $N$ is first split into a series of $S$ chunks, each of a fixed length $T$.
The model applies a chunk transformer encoder within each segment, consisting of a self-attention mechanism (\Cref{eq:self_att}) followed by a feed-forward neural network.
However, the self-attention mechanism in each chunk also attends to the hidden states from the previous chunk, using a recurrence mechanism, which allows the memory to scale linearly with respect to $N$.
This recurrent mechanism helps the model to attend to past predictions and enlarge its context.
The keys $\mathbf{K}_{c} \in \mathbb{R}^{d \times T }$ and values $\mathbf{V}_{c} \in \mathbb{R}^{d \times T}$ for chunk $c$, are concatenated with their previous hidden outputs as:

\begin{align}
    \Tilde{\mathbf{K}}_{c} = [sg(\mathbf{K}_{c-1}) \mathbin\Vert \mathbf{K}_c], \Tilde{\mathbf{V}}_{c} = [sg(\mathbf{V}_{c-1}) \mathbin\Vert \mathbf{V}_c], 
\end{align}

where $sg(.)$ is a stop-gradient operation and $[\cdot\mathbin\Vert\cdot]$ is the concatenation operation over the (time) dimension $T$.
The stop-gradient operation is necessary to avoid vanishing/exploding gradients and this operation is done for each layer $l=1,\ldots,L$.
Note the length $T_c$ of $\mathbf{K}_{c-1} \in \mathbb{R}^{d \times T_c}$ might differ from $\mathbf{K}_c$. 
Throughout our research, we conducted experiments with $T_c > T$; however, we observed no performance improvements. 
This lack of improvement can be attributed to the spontaneous characteristics of audio signals in Audioset. For instance, consecutive chunks may contain the same sound event, whereas chunks that are distant from each other may not share common sound events.


\section{Experiments}

\begin{figure}[htbp]
    \centering
    \includegraphics[width=\linewidth]{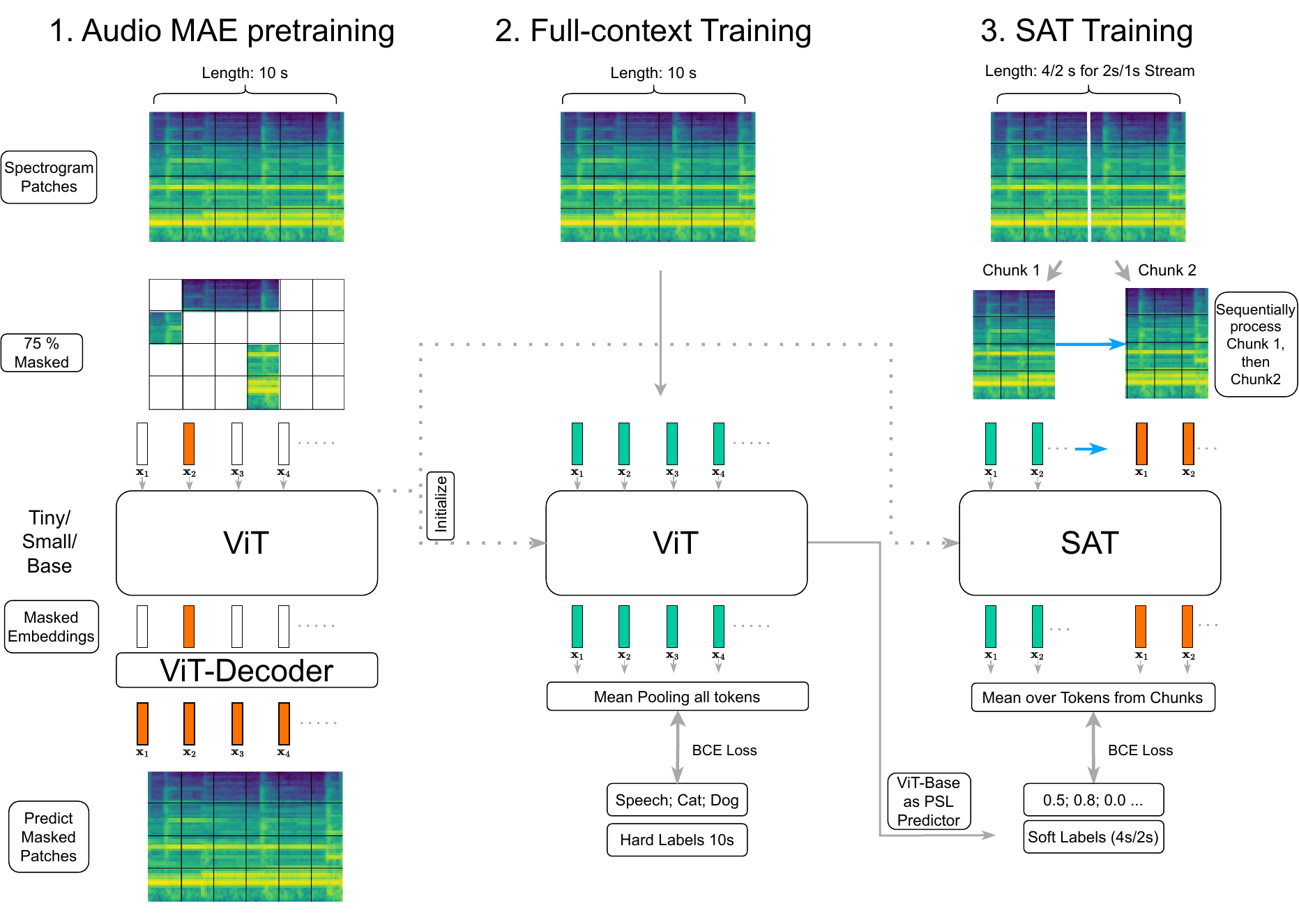}
    \caption{The proposed training pipeline consists of three stages. First, pretraining using masked auto-encoders (MAE), second we use standard full-context training (10s clips) and third, our best model (ViT-B) is used to predict labels on a fine scale for SAT training.}
    \label{fig:framework}
\end{figure}

\subsection{Datasets}

Our training and evaluation dataset is Audioset~\cite{gemmeke2017audio}, which mainly contains 10-second long audio clips.
We collected 1,904,746 training samples (5200 h) and 18,299 evaluation samples (50 h) sampled at 16 kHz.
While our baseline models are trained and evaluated on the full context of an audio clip (10s), the proposed streaming transformers are mismatched between training (1s and 2s) and evaluation (10s).
Thus, during evaluation, for clips longer than the chunk length, we split the input into equally sized chunks, then feed these chunks into the model and average all output scores.

\subsection{Training Pipeline}

Our training pipeline can be summarized into three steps (see \Cref{fig:framework}).
To achieve competitive performance, we first pretrain all our transformer models via the masked autoencoder (MAE) paradigm~\cite{he2022masked}.
Second, we finetune the pretrained transformer model on Audioset using full-context training (10s).
Third, pseudo strong labels (PSL)~\cite{dinkel2022pseudo} are used to predict soft labels on a target length depending on the required delay and train each SAT model with these (soft) labels.

\subsection{Setup}
\label{ssec:setup}

The present study employs three widely recognized ViT-based architectures, namely ViT-Tiny (ViT-T), ViT-Small (ViT-S), and ViT-Base (ViT-B)~\cite{Dosovitskiy_ViT}. 
Each of these models comprises a $L=12$ layer transformer with an embedding dimension $d$ and a number of heads $N_h$, $(d,N_h)$ set to $(192, 3),(384, 6)$ and $(768, 12)$ for ViT-T/S/B, respectively.
This study aims to explore the practical application of attention-based models, prioritizing factors such as inference speed and memory usage over marginal gains in performance and thus we optimize the training pipeline to use as few resources as possible. 
In contrast to previous works~\cite{gong21b_interspeech,gong2022cmkd,chen2022beats,baade22_interspeech,koutini22_interspeech}, we use an initial batch-norm to normalize each Mel-filterbank independently.
Furthermore, we employ 64-dimensional banks at a 16 kHz sampling rate, extracted within a 32 ms window and a 10 ms hop instead of the conventional 128-dimensional banks at 32 kHz. 
This significantly improves inference speed and lowers memory usage.
Also, we train with an 8-bit Adam optimizer~\cite{dettmers2022optimizers} to further conserve memory. 
As a result, the ViT-B model only requires 6 GB of memory during training with a standard batch size of 32. 
Therefore, it is feasible to train our models on a single graphics processing unit (GPU) without the need for large computation clusters.
The neural network back-end is implemented in Pytorch~\cite{PaszkePytorch} and the source code with pretrained checkpoints is publicly available\footnote{\url{https://github.com/RicherMans/SAT}}.
We further provide details regarding each stage of the training process.

\paragraph*{Self-supervised pretraining}

We pretrain three transformers, one for each size using a vanilla MAE~\cite{he2022masked,xu2022masked} approach.
In the MAE framework, 75\% of the patches (see \Cref{eq:patchification}) from a spectrogram $\mathbf{S}$ are removed and forwarded through a ViT encoder.
Then a decoder is tasked to reconstruct the masked patches.
We append a ViT decoder with $d=512, N_h = 8$, where $L=8$ for ViT-S/B and $L=4$ for ViT-T.
After training, the decoder is removed and only the encoder is kept, which serves as the pretrained checkpoint for full-context and SAT training.
We train on a single GPU with a batch size of 64, a starting learning rate of 2e-4, a linear warmup of 3 epochs and a final learning rate of 2e-5 using a half-cycle cosine decay~\cite{loshchilov2016sgdr}.
The ViT-B model is pretrained for 40 epochs, while the ViT-T/S models run for 50 epochs.

\paragraph*{Training of baseline Audioset models}

For all models, we use time- and frequency-independent absolute learnable position embeddings~\cite{koutini22_interspeech}.
Further, we employ random shifting, volume gain, and polarity inversion as augmentation methods in the waveform domain and Specaugment~\cite{park19e_interspeech}, which masks 192 consecutive time-frames and 24 frequency bins, in the spectrogram domain.
Mixup~\cite{zhang2018mixup} in the waveform domain is also employed with $\alpha=0.3$. 
Importance sampling based on the label frequency is applied to counter the long tail of infrequent classes, where one epoch in our training scheme consists of sampling 200 k audio clips without replacement.
Further, we use patch-out~\cite{koutini22_interspeech} where we randomly drop 25\% of the time- and 25\% of the frequency patches and use stochastic depth regularization~\cite{huang2016deep} with a probability of 0.1.
The training objective is binary cross entropy (BCE) and runs for at most 180 epochs with learning rates of 5e-4/4e-4/1e-5 and a warmup of 60/55/10 epochs for ViT-T/S/B, respectively.
For all experiments, we average the top-4 models with the highest mAP, the main evaluation metric.

\subsection{Training of streaming transformers}

SAT models' recurrent dependency on previous chunks during training requires looping over sequential segments, decreasing efficiency.
We use PSL~\cite{dinkel2022pseudo}, where our best model (ViT-B) is tasked to predict fine labels on a 4/2s scale for the 2/1s delay streaming models respectively, such that SAT has access to a single past chunk.
The SAT models are trained without augmentation and mixup, by randomly sampling 200 k crops from the dataset each epoch.
Training runs for at most 250 epochs, with identical learning rate schedules as their full-context ViT counterparts.
Evaluation on Audioset is done by feeding each model chunks of its respective delay and averaging these scores across each 10s clip.

\section{Results}

\subsection{Baseline results for full-context audio tagging}

Our primary objective is to validate our proposed full-context models by comparing them with other transformer-based works from the literature.
We also provide flops, and peak memory requirements (without considering the model parameters) for running inference on a single sample.
The results obtained from our training pipeline, as shown in \Cref{tab:baseline_results_shuabang}, indicate that our models, ViT-T (44.2) and ViT-S (45.7), achieve SOTA results using significantly fewer tokens as compared to previous works. 
Even though our proposed ViT-B (47.4) underperforms against BEATs~\cite{chen2022beats} (47.9), it also requires three times less memory and approximately half of the computational resources. 
Additionally, it is worth mentioning that while HTS-AT~\cite{chen2022hts} consumes less memory and requires fewer Gflops, the model is strictly optimized for offline usage, which we will further elaborate on in \Cref{ssec:streaming}.

\begin{table}[tb]
    \centering
    \begin{tabular}{l|rrrrr}
        \toprule
        Model & Size & \#Token &  $\hat{M}$ & Gflops & mAP \\
        \midrule
        PaSST-Tiny~\cite{chong2022masked} & \multirow{3}{*}{5.6} & 496 & 120 M & 5.4 & 39.7 \\
        MS-Tiny~\cite{chong2022masked} &  & 496 & 120 M & 5.4 & 40.3 \\
        Ours (ViT-T) &  & \textbf{256} & \textbf{52 M} & \textbf{2.7} & \textbf{44.2} \\ 
        \midrule
        PaSST-S~\cite{chong2022masked} & \multirow{4}{*}{22} & 496 & 279 M & 21 & 43.1 \\
        MS-Small~\cite{chong2022masked} &  & 496 & 279 M  & 21 & 44.2 \\
        AMAE-S$^{\dagger}$~\cite{xu2022masked} & & 512 & 291 M & 22 & 45.0 \\
        Ours (ViT-S) &  & \textbf{256} & \textbf{105 M} & \textbf{10}  &\textbf{45.7}\\
        \midrule
        AST$^{\dagger}$~\cite{gong21b_interspeech} & 86 & 1212 & 2.2 G & 202 & 45.9 \\
        KD-AST$^{\dagger}$~\cite{gong2022cmkd} & 86 & 1212 & 2.2 G & 202 & 47.1\\
        MS-Base~\cite{chong2022masked} & 86 & 496 & 575 M & 83 & 47.1 \\
        PaSST~\cite{koutini22_interspeech} & 86 & 512 & 562 M & 85 & 47.1 \\
        AMAE-B$^{\dagger}$~\cite{xu2022masked} & 86 & 512 & 562 M & 85 &  47.3 \\
        AMAE-L$^{\dagger}$~\cite{xu2022masked} & 304 & 512 & 1.5 G & 300 & 47.4 \\
        HTS-AT~\cite{chen2022hts} & 31 & 1024 & \textbf{171 M} & \textbf{14} & 47.1 \\
        BEATs$^{\dagger}$~\cite{chen2022beats} & 90 & 496 & 620 M & 100 & \textbf{47.9}\tablefootnote{Our own evaluation of the public checkpoint reproduced 45.5. AST and HTS-AT are consistent with their reported results.} \\
        Ours (ViT-B) & 86 & \textbf{256} & 210 M & {42} & {47.4} \\
        \bottomrule
        \end{tabular}
    \caption{Baseline results of ViT-based AT models with pretraining compared to other works. 
    Input length in number of tokens is denoted by \#token, and model size is in millions. 
    $\hat{M}$ represents peak memory in bytes during a forward pass (excluding model parameters). Models with $^\dagger$ require multiple GPUs. The best result for each category is in bold.}
    \label{tab:baseline_results_shuabang}
\end{table}

\vspace{-1mm}
\subsection{Streaming Audio Transformers}
\label{ssec:streaming}

The results are displayed in \Cref{tab:main_results}, where we compare them against our own baseline models as well as SOTA approaches such as AST~\cite{gong21b_interspeech}, BEATs~\cite{chen2022beats} and HTS-AT~\cite{chen2022hts}.
For the evaluation of these SOTA approaches, we utilize their respective publicly available checkpoints with a delay of 2s. This means that we provide each model with a chunk of length 2s and average the scores, which aligns with how we evaluate our streaming models.
Results from the experiment show that all models perform worse when evaluated with a 2s delay compared to full-context evaluation (see \Cref{tab:baseline_results_shuabang}). 
This is primarily due to the limitation of accessing only a single chunk at a time, which restricts the models' ability to utilize future context, which is necessary for well-performing \textit{offline} AT models.
However, while AST (46.9 $\rightarrow$ 39.7), BEATs (47.9 $\rightarrow$ 38.7), and HTS-AT (47.1 $\rightarrow$ 5.2) all showed a significant drop in performance when evaluated with a delay of 2s compared to their respective full-context counterpart, our most efficient SAT-T model outperformed all baselines with significantly lower memory and flops requirements. 
This suggests that the other approaches may be tailored towards their target duration of 10s and struggle to be effective for real-world online evaluation.
Our best SAT-B model with a 2s delay achieves an mAP of 45.1, which is 0.8 mAP points lower than full-context AST but uses 1.6\% of AST's memory and 3.7\% Gflops, while also having a significantly shorter delay of 2s.
Further decreasing the delay to 1s decreases performance across all evaluated models, even though our proposed SAT-T/S/B models still significantly outperform the baselines in this condition.

\begin{table}[tb]
    \centering
    \begin{tabular}{ll|rrrrr}
    \toprule
         & Model  & Strm? & \#Token & $\hat{M}$ & GFlops & mAP   \\
         \midrule
          \parbox[t]{1mm}{\multirow{9}{*}{\rotatebox[origin=c]{90}{2s delay}}}  & ViT-T  & \multirow{6}{*}{\xmark}  & 48 & 7.6 M & 0.5  & 39.1 \\
         & ViT-S  &  & 48 & 15 M & 2.1 & 40.9 \\
         & ViT-B &  & 48 & 30 M & 8.2 & 41.5 \\
         & AST$^{\star}$  &  & 1212 & 2.2 G & 202 & 39.7 \\
         & BEATs$^{\star}$  &  & 96 & 83 M & 17.8  & 38.7\\
         & HTS-AT$^{\star}$  &    & 1024 & 171 M  & 42 & 5.2  \\
         \cline{2-7}

         & SAT-T & \multirow{3}{*}{\checkmark} & \multirow{3}{*}{48/48} & {9 M} & {0.5} & 43.3 \\
         & SAT-S &   &   & 18 M & 2.1 & {43.4} \\
         & SAT-B &   &  & 36 M & 8.2 & {45.1} \\
         \midrule
         \midrule
          \parbox[t]{1mm}{\multirow{9}{*}{\rotatebox[origin=c]{90}{1s delay}}}  & ViT-T  & \multirow{6}{*}{\xmark} & 24 & 3.8 M & 0.3 & 33.0\\
          & ViT-S &  & 24 &  7.5 M & 1.1 & 34.9 \\
          & ViT-B  &  & 24 & 14 M & 4.1 & 34.2 \\
          & AST$^{\star}$  &   & 1212 & 2.2 G & 202 & 36.6 \\
         & BEATs$^{\star}$  &    & 48 & 83 M & 17.8  & 35.2\\
         & HTS-AT$^{\star}$  &     & 1024 & 171 M  & 42 & 2.4  \\
         \cline{2-7}
         & SAT-T & \multirow{3}{*}{\checkmark} & \multirow{3}{*}{24/24} & 4.3 M & 0.3 & 40.1  \\
         & SAT-S &   &  & 9 M & 1.1  & 40.2 \\
         & SAT-B &   &  & 16 M & 4.1 & 41.4\\
         \bottomrule
    \end{tabular}
    \caption{Streaming Transformer results on Audioset using evaluation with a 2s and 1s delay. When streaming is used, \#Token refers to $T$/$T_c$. ``Strm?'' indicates whether the model is streaming. $\hat{M}$ refers to the peak memory consumed (in bytes) during a forward pass of a single sample (with cache). $^{\star}$ denotes evaluation from a public checkpoint. HTS-AT and AST pad their input to 10s, thus Gflops and $\hat{M}$ are unaffected.}
    \label{tab:main_results}
\end{table}

\subsection{Segment-level evaluation}
To further validate the capability of SAT in accurately predicting chunk-level tags, we conducted evaluations using the strongly labeled Audioset evaluation set~\cite{hershey2021benefit}, which contains 381 matching labels with the weakly labeled Audioset used in this study.
However, it is important to note that the labeling scheme differs between the two sets. For instance, the labeling scheme in the strongly labeled Audioset assigns labels such as ``Speech'' to either ``Male'' or ``Female,'' resulting in significant performance discrepancies.
We present the Segment-F1 (Seg-F1) and event onset-F1 scores~\cite{Mesaros2016MetricsFP}, obtained by thresholding each model's output with a value of 0.5 at each respective chunk length. 
As the results in \Cref{tab:seg_f1_results} indicate, our proposed SAT approach consistently outperforms other powerful transformer models in terms of segment-level performance on the strongly labeled Audioset.

\begin{table}[htbp]
    \centering
    \begin{tabular}{ll|rr|rr}
    
        \toprule
         &  & \multicolumn{2}{c}{2s} & \multicolumn{2}{c}{1s} \\
         Model  & Strm? & Seg-F1 & Onset-F1 & Seg-F1 &  Onset-F1 \\
         \midrule
         ViT-T  & \multirow{5}{*}{\xmark}  &  18.2 & 6.2 & 11.1 &  6.2 \\
         ViT-S  &  &  19.0 & 6.4  & 11.7 &  6.4 \\
         ViT-B &  &  18.8 & 6.2 & 10.2 & 6.2 \\
         AST  &  & 13.7 & 4.7 & 10.5 & 4.7 \\
         BEATs & &  16.1 & 5.7 & 11.0  & 5.7 \\
         \hline
         SAT-T & \multirow{3}{*}{\checkmark} & 28.9 & 8.6 & 24.2  & 7.3 \\
         SAT-S &   & {30.6} & 9.0 &  25.3 & 7.1\\
         SAT-B &   & {30.1} & 8.4 &  22.3 & 8.3 \\
         \bottomrule
    \end{tabular}
    \caption{Results for the segment and onset-based F1-score evaluation on the strongly labelled Audioset evaluation set using a threshold of 0.5 for 2s and 1s input segments. Models have been trained on the weakly labeled Audioset.}
    \label{tab:seg_f1_results}
\end{table}

\subsection{Quantitative Analysis}
\label{ssec:quantative}

A particularly useful application for streaming transformers is the (continuous) detection of long-duration sound events.
However, to the best of our knowledge, there exists no public dataset that contains long audio samples that can simulate real-world audio streams.
\begin{figure}[tb]
    \centering
    \includegraphics[width=1.06\linewidth]{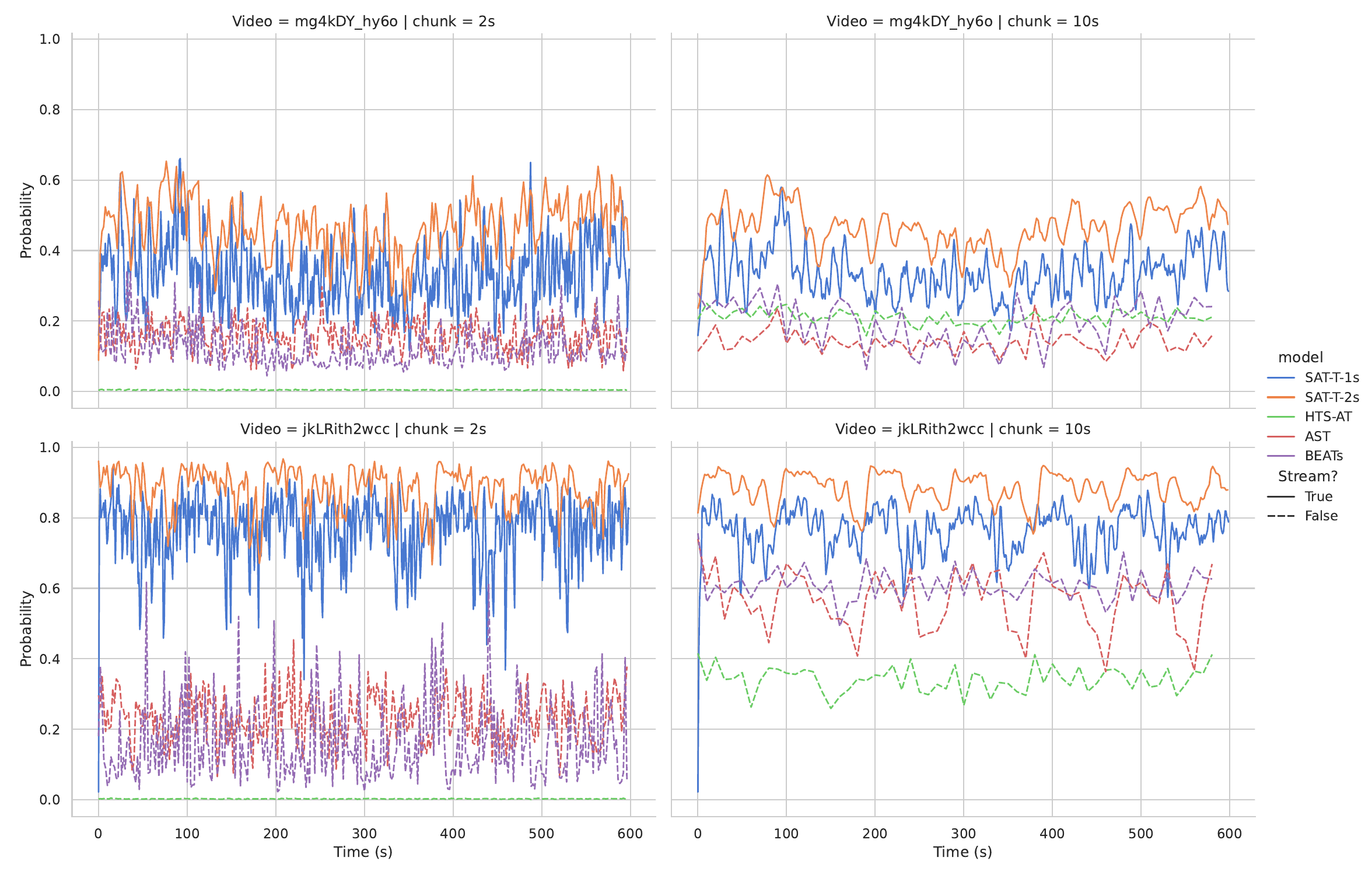}
    \caption{Comparison of output probability scores between the baselines against the proposed SAT-T for a 10-minute long sound of water event. Samples (Top: \url{mg4kDY_hy6o}, Bottom: \url{jkLRith2wcc}) were taken from Youtube and evaluated using 2s (left) and 10s (right) chunks. Best viewed in color. }
    \label{fig:quantative_streams}
\end{figure}
Thus, we collected two 10-minute water sound samples from YouTube\footnote{Accessible via \url{www.youtube.com/watch?v=VIDEO_ID}} and evaluated different AT models as depicted in \Cref{fig:quantative_streams}. 
Two sets of experiments were conducted, one at a full-context (10s) scale, matching the delay of standard AT approaches, and one with limited-context (2s) inputs matching the resolution of our proposed SAT.
Note that these two samples are monotonous and repetitive; in other words, they should be easily detected by a potent AT model with a consistently high probability over a prolonged amount of time.
We compared our smallest model, SAT-T, with BEATs, HTS-AT, and AST, and found that SAT-T consistently outperformed the competitors in predicting the presence of water with high confidence and lower delay for all experimented settings. 
Further, due to its strict optimization to a 10s delay, HTS-AT is incapable to predict any presence of water, which is in line with the observations in~\Cref{tab:main_results}.

\section{Conclusion}

This paper presents streaming transformers for AT, which have lower memory usage, faster response time, and can capture long-range sound events. 
The paper proposes three ViT-based models (SAT-T/S/B) with two streaming configurations for short-delay AT. 
The best model, SAT-B, achieves an mAP of 45.1 with a 2s delay, using 8.2 Gflops and 36 MB of memory during inference. 
The paper also demonstrates through a simulation that the proposed streaming transformers outperform other state-of-the-art methods in terms of long-term consistency.



\bibliographystyle{IEEEtran}
\bibliography{mybib}

\end{document}